\documentclass[preprint]{aastex}

\usepackage{mathtext,bbm,amsmath,amsfonts,amssymb,indentfirst,syntonly,graphicx}
\usepackage[english]{babel}
\usepackage{slashbox}
\usepackage{calc}
\usepackage{tikz}
\usepackage{latexsym,graphicx,bbm}

\newcommand{\beq}{\begin{eqnarray}}
\newcommand{\eeq}{\end{eqnarray}}

\begin{document}

\title{Gamma-ray burst polarization via Compton scattering process}
\author{Zhe Chang\altaffilmark{1,2}, Hai-Nan Lin\altaffilmark{1,*}, Yunguo Jiang\altaffilmark{3,4}}
\affil{\altaffilmark{1}Institute of High Energy Physics\\Chinese Academy of Sciences, 100049 Beijing, China}
\affil{\altaffilmark{2}Theoretical Physics Center for Science Facilities\\Chinese Academy of Sciences, 100049 Beijing, China}
\affil{\altaffilmark{3}School of Space Science and Physics \\Shandong University at Weihai, 264209 Weihai, China}
\affil{\altaffilmark{4}Shandong Provincial Key Laboratory of Optical Astronomy \\ and Solar-Terrestrial Environment, 264209 Weihai, China}
\altaffiltext{*}{linhn@ihep.ac.cn}

\begin{abstract}
Synchrotron radiation and Compton scattering are widely accepted as the most likely emission mechanisms of some astrophysical phenomena, such as gamma-ray bursts (GRBs) and active galactic nuclei (AGNs). The measurement on polarization of photons provides a useful tool to distinguish different emission mechanisms and structures of the emission region. Based on the differential cross section of a polarized photon scattered by an unpolarized electron of any initial momentum, we derive analytical formula of polarization for beamed photons scattered by isotropic electrons with a power law distribution. Numerical calculations are carried out in four special cases: electrons at rest, Thomson limit, head-on collision and monochromatic electrons. It is found that the maximum polarization can be as high as $100\%$ for low energy photons, if the electrons are at rest. Although polarization is highly suppressed due to the isotropic electrons, a maximum value of $\sim 10\% \-- 20\%$ can still be achieved. Compton scattering process can be used to explain the polarization of GRB 041219A and GRB 100826A.
\end{abstract}

\section{Introduction}

Gamma-ray bursts (GRBs) are the most energetic explosions in the universe. The isotropic equivalent energy reaches $\sim 10^{54}$ ergs for the brightest bursts. Since their discovery in the 1960s, after decades of researches, a lot of progresses have been made in both the observational and theoretical aspects. The spectrum of prompt emission can often be fitted well by a broken power law, namely, the Band function \citep{Band:1993}. The $\nu F_{\nu}$ spectrum generally peaks at $0.1\--1$ MeV. Synchrotron is widely accepted as one of the most promising emission mechanisms \citep{Meszaros1994,Tavani:1996,Dermer:1999,Lloyd:2000}. Since the launch of {\it FERMI} satellite, GeV photons have been observed in the prompt emission of some GRBs \citep{Abdo:2009,Abdo:2009yy,Abdo:2009mm,Ackermann:2011}. An interesting feature is that GeV photons often arrive seconds later relative to MeV photons. If GeV photons originate from synchrotron, the Lorentz factor of the electrons should be very large and the magnetic field must be extremely strong, which are hardly realized in the astrophysical conditions. \citet{Meszaros2011} showed that the magnetic-dominated jet model can naturally explain the delayed arrival of high energy photons. According to this model, a jet dominated by Poynting flux but contaminated by baryons, emits from the central engine with a large Lorentz factor $\Gamma$. MeV photons are produced by synchrotron of electrons, while GeV photons are produced by neutron-proton collisions or Compton scattering process. The optical depth of GeV photons is larger than that of MeV photons. Thus, GeV photons emit at a larger radius where the optical depth becomes small enough. The time delay between high and low energy photons gives the constraint on jet Lorentz factor $\Gamma$. For long GRBs $\Gamma\sim 200$, and for short GRBs it is much larger \citep{Chang:2012}. The magnetic-dominated jet model can produce a various type of spectra \citep{Veres:2012sb}.

Besides the spectra, polarization measurement of photons provides another direct insight into the nature of GRBs. In spite of many controversies exist, polarization has been observed in the prompt or afterglow phase of some GRBs (for example, see Table 1 of \citet{Chang:2013}). \citet{Coburn:2003} reported a polarization of $80\%\pm 20\%$ in the prompt phase of GRB 021606. \citet{Rutledge:2004} re-checked the data but no significant evidence for polarization was found. \citet{Wigger:2004} used a novel method to re-analyze the data and found a linear polarization of $41_{-44}^{+59}\%$, which shows that the data is too poor to make a convincing conclusion. \citet{Kalemci:2007} analyzed the data of the prompt emission of GRB 041219A in the energy band $100\-- 350$ keV, and found a linear polarization $98\%\pm 33\%$, although the instrumental systematics cannot be ruled out. \citet{McGlynn:2007} investigated the same GRB in three different energy bands, and found that the polarization tends to decrease as the energy increases. \citet{Gotz:2009} re-examined the data and found a variable degree of polarization ranging from less than $4\%$ over the first peak to $43\%\pm 25\%$ for the whole second peak. Similar debates occur in the afterglow phase. However, the polarization in the GRB afterglow phase seems to be smaller than that in the prompt phase. The polarization of GRB afterglow is generally less than $10\%$ \citep{Covino:1999,Hjorth:1999,Wijers:1999,Bersier:2003}.

Many theoretical works have been devoted to studying the polarization of GRBs. It is well known that the maximum polarization of synchrotron is $\Pi_{\rm max}=(p+1)/(p+7/3)$, if the electrons are isotropic with a power-law index $p$ and the magnetic field is uniform globally \citep{Rybicki:1979}. For a typical value of $p\sim 3$, one has $\Pi_{\rm max}\sim 75\%$. If the magnetic field contains $N$ uniform patches, the net polarization is about $\Pi_{\rm max}/\sqrt{N}$ \citep{Gruzinov:1999a}. A randomly oriented magnetic field may also produce significant polarization if the line-of-sight is close to the jet edge \citep{Waxman:2003}. However, this occurs by chance only if the jet opening angle is very small. \citet{Lazzati:2004} investigated the Compton drag model and showed that the net polarization can be large if certain geometrical conditions are realized. \citet{Toma:2009} performed the Monte Carlo simulation to calculate the GRB polarization in three different models: the synchrotron model with a globally ordered magnetic field, the synchrotron model with a small-scale random magnetic field, and the Compton drag model. They found that the Compton drag model is favored if the polarization is larger than $80\%$. \citet{Lundman:2013qba} investigated the polarization properties of photospheric emission originating from highly relativistic jet, and found that the polarization degree can reach $\sim 40\%$ in particular situation. \citet{Mao:2013gha} studied the jitter radiation and concluded that high degree of polarization can be achieved. It was also showed that photons scattered off highly relativistic, baryon-rich materials can produce significant polarization \citep{Eichler:2003,Eichler:2004aa,Eichler:2004bb}.

In a very recent paper, \citet{Chang:2013yma} presented an analytical formalism for the polarization of beamed photons scattered by isotropic electrons. The polarization of incident photons can be any value and the energy of incident electrons can be any distribution. It was showed that the photon-electron scattering may produce significant polarization even when the electrons have an isotropic momentum distribution, regardless that the electrons have thermal or nonthermal (such as power-law distribution) energy spectra. However, the formulae in that paper were given in the jet comoving frame. In addition, the polarization was expressed as a function of incident photon energy, which is not an observable quantity. In this paper, we try to transform the formulae into the observer frame. Furthermore, the polarization is expressed as a function of the energy of scattered photon, which can be detected directly. This makes the formulae more convenient to be used in astrophysical processes.

The rest of the paper is arranged as follows. In Section \ref{sec:general-formulae}, we introduce the kinematics of the general Compton scattering process, obtain the formulae of Stokes parameters for scattered photons, and derive final polarization analytically as a function of photon energy and viewing angle. In Section \ref{sec:result}, we carry out numerical calculations of polarization in four special cases: (1) electrons at rest, (2) Thomson limit, (3) head-on collision, (4) monochromatic electrons. Although these four cases may be far away from the actual astrophysical processes, they show the main properties of photon polarization via Compton scattering process. In section \ref{sec:GRB-cases}, we test the validity of these formulae in explaining the observational data on two specific GRBs, i.e., GRB 041219A and GRB 100826A. Finally, discussions and conclusions are given in section \ref{sec:conclusions}.

\section{General formulae of photon polarization via Compton scattering process}\label{sec:general-formulae}

\begin{figure}
\centering
  \plotone{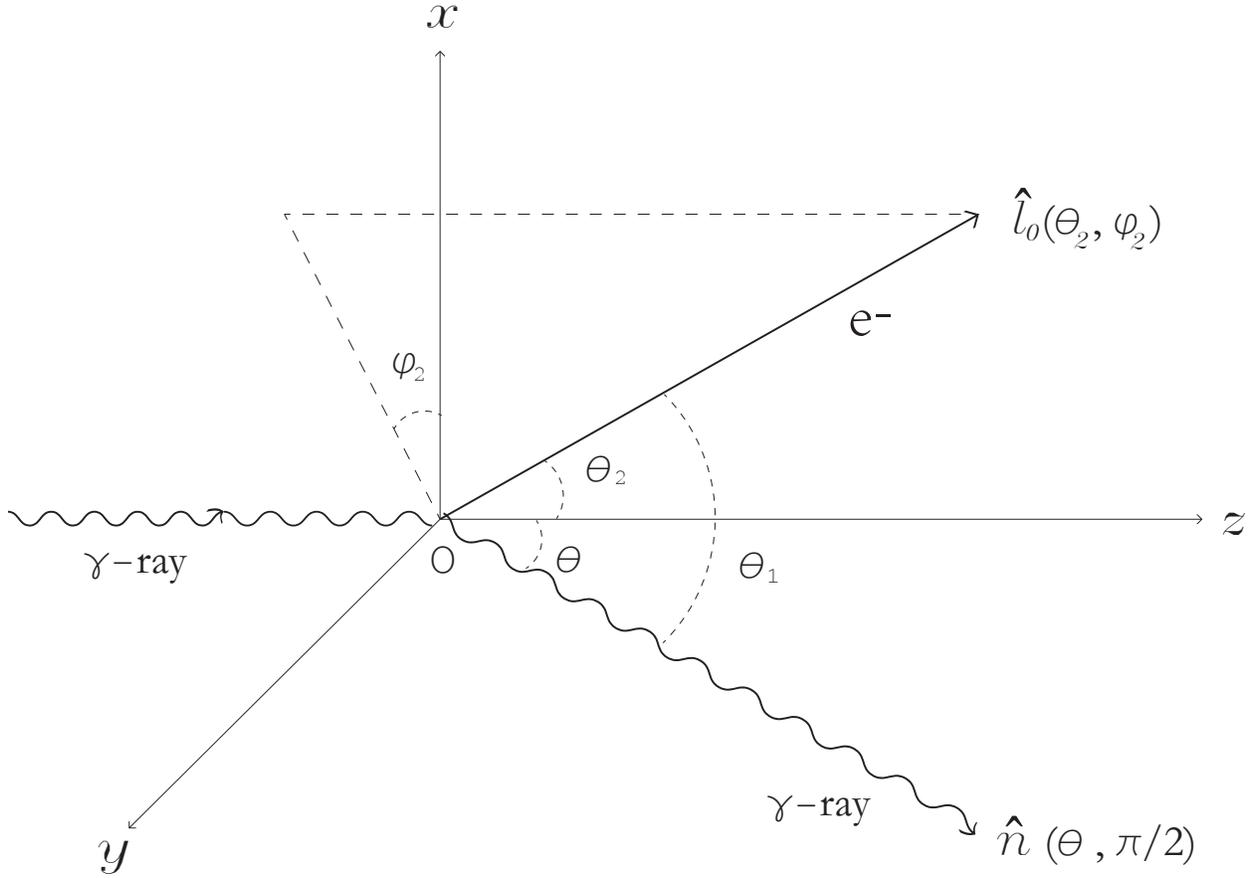}
  \caption{\small{Schematic representation of Compton scattering process in the jet comoving frame. The incident photon goes along the positive $z$-axis, scattered by an electron  at point $O$, then moves along the line-of-sight $\hat{{\bf n}}$. The initial electron is injected along $\hat{{\bf l}}_0$ direction, and the moving direction of the scattered electron is ignored here. We choose a Cartesian coordinate system such that the $y$-axis is in the scattering plane. The polar and azimuthal angles of $\hat{{\bf l}}_0$ and $\hat{{\bf n}}$ are ($\theta_2, \varphi_2$) and ($\theta, \pi/2$), respectively. The angle between $\hat{{\bf l}}_0$ and $\hat{{\bf n}}$ is denoted by $\theta_1$.}}\label{fig:geo}
\end{figure}

The general process of Compton scattering is depicted in Fig.\ref{fig:geo}. Consider a photon with energy $\varepsilon_0$ moves along the $z$-axis, then collides with an electron at point $O$. After scattering, the photon goes towards $\hat{{\bf n}}$ direction\footnote{We take the convention that a vector with a hat denotes the unit vector along that direction.}. Define a Cartesian coordinate system such that the $y$-axis is in the scattering plane (the plane which contains $\hat{{\bf z}}$ and $\hat{{\bf n}}$), and the $xyz$ axes form a right-handed set. The injected electron has an arbitrary momentum ${\bf p}_0= \gamma \beta m_e c \hat{{\bf l}}_0$, where $\gamma$ is the Lorentz factor of the electron, $\beta=|{\bf v}|/c$ is its velocity in unit of light speed, and $\hat{{\bf l}}_0$ is the moving direction of the incident electron. The parameter space of the incident electron can be completely represented by coordinates ($\gamma, \theta_2, \varphi_2)$, where $\theta_2\in [0,\pi]$ and $\varphi_2\in [0,2\pi]$ are the polar and azimuthal angles of the injected electron, respectively.

The energy of scattered photon can be deduced from the conservation of energy and momentum. It reads \citep{Akhiezer:1965}
\begin{equation} \label{eq:epspr}
 \varepsilon_1=\frac{\varepsilon_0(1-\beta \cos\theta_2)}{\frac{\varepsilon_0}{\gamma m_e c^2}(1-\cos\theta)+(1-\beta\cos\theta_1)},
\end{equation}
where $\cos\theta_2=\hat{{\bf z}}\cdot\hat{{\bf l}}_0$,  $\cos\theta=\hat{{\bf z}}\cdot\hat{{\bf n}}$, and $\cos\theta_1=\hat{{\bf l}}_0\cdot\hat{{\bf n}}$. From the geometrical considerations, we have
\begin{equation}\label{eq:angrel}
\cos \theta_1 = \cos \theta \cos \theta_2 + \sin \theta \sin \theta_2 \sin \varphi_2.
\end{equation}
One can also obtain the Lorentz factor and moving direction of the scattered electron. However, we are only interested in the scattered photon, the scattered electron is ignored here.

The polarization of a photon can be conveniently described by Stokes parameters $\xi_i$ ($i=1,2,3$) \citep{Berest:1982}. They are real numbers and are defined with respect to the $xyz$ axes. The positive (negative) $\xi_3$ describes photon linearly polarized along the $x$ ($y$) axis. The parameter $\xi_1$ means the linear polarization along the directions with azimuthal angles $\pm\pi/4$ relative to the $x$-axis in the $xy$ plane. The parameter $\xi_2$ represents left-handed or right-handed circular polarization. In terms of Stokes parameters, the degree of polarization can be written as
\begin{equation}\label{eq:pi}
 \Pi_0 = \sqrt{\xi_1^2+\xi_2^2+\xi_3^2}.
\end{equation}
The Stokes parameters satisfy the condition $0\leq \xi_1^2+\xi_2^2+\xi_3^2 \leq 1$. If $\Pi_0=0$, the photon is unpolarized. If $\Pi_0=1$, the photon is completely polarized. In the $0< \Pi_0 < 1$ case, we call that the photon is partially polarized.

In this paper, we are only interested in linear polarization. The circular polarization is generally very small, especially in afterglow phase \citep{Matsumiya:2003a}. For simplicity, we ignore the circular polarization and set $\xi_2=0$ in the following discussions. The polarization direction of a photon can be denoted by
\begin{equation}
  \tan2\chi_0=\frac{\xi_1}{\xi_3},
\end{equation}
where $\chi_0$ is the angle between the polarization direction and the $x$-axis. If $\xi_1=0$, i.e., $\chi_0=0$ or $\pi/2$, the polarization is along the $x$-axis or $y$-axis, respectively. On the other hand, if $\xi_3=0$, i.e., $\chi_0=\pm\pi/4$, the polarization is along the directions with azimuthal angles $\pm\pi/4$ relative to the $x$-axis in the $xy$ plane.

The differential cross section for the scattering of a polarized photon by an unpolarized electron is written as \citep{Berest:1982,Chang:2013yma}
\begin{equation}\label{eq:cs1}
  d \sigma = \frac{1}{4} r_e^2 d \Omega \left(\frac{\varepsilon_1}{\varepsilon_0}\right)^2  \bigg[ F_0 +F_3(\xi_3+\xi'_3) + F_{11} \xi_1 \xi_1' +F_{22} \xi_2\xi'_2+F_{33} \xi_3\xi'_3\bigg],
\end{equation}
where $d\Omega=\sin \theta d \theta d \varphi$ is the solid angle, $r_e=e^2/m_ec^2$ is the classical electron radius, and
\begin{eqnarray}\label{eq:fs}
 \begin{cases}
  F_0 \equiv \frac{\varepsilon_1}{\varepsilon_0}+\frac{\varepsilon_0}{\varepsilon_1}-\sin^2 \theta,\\
  F_3 \equiv -(A^2+A) \Sigma,\\
  F_{11}\equiv (A+\frac{1}{2}) \Sigma,\\
  F_{22}\equiv \frac{1}{4}(1+2A)B \Sigma,\\
  F_{33}\equiv (A^2+A+\frac{1}{2}) \Sigma,
 \end{cases}
\end{eqnarray}
\begin{equation}\label{eq:ab}
  A \equiv \frac{1}{x}-\frac{1}{y}, \quad B\equiv \frac{x}{y}+\frac{y}{x}, \quad \Sigma\equiv\frac{4}{\gamma^2(1-\beta\cos\theta_2)^2}\left(1-\frac{\beta\sin\theta_2\sin\varphi_2}{1-\beta\cos\theta_2} \tan\frac{\theta}{2}\right),
\end{equation}
\begin{equation} \label{eq:xy}
  x=\frac{2\gamma \varepsilon_0}{m_e c^2}(1-\beta \cos \theta_2), \quad
  y=\frac{2\gamma \varepsilon_1}{m_e c^2}(1-\beta \cos \theta_1).
\end{equation}
The Stokes parameters $\xi_i$ and $\xi'_i$ describe the polarization of the incident and scattered photons, respectively. It should be noted that $\xi'_i$ are defined in a new coordinate system $O'x'y'z'$, whose $x'$-axis is parallel to the $x$-axis, $z'$ axis is along the direction $\hat{{\bf n}}$, and $\hat{y'}=\hat{z'}\times\hat{x'}$. In other words, the $O'x'y'z'$ system is the $Oxyz$ system rotating an angle $\theta$ relative to the $x$-axis.  This makes sure that the polarization direction of the scattered photon is still perpendicular to its wave vector.

The Stokes parameters of the secondary photon are denoted by $\xi^{\rm f}_i$, which are equal to the ratios of the coefficients of  $\xi'_i$ to the terms independent of $\xi'_i$ \citep{Berest:1982}, i.e.,
\begin{equation}\label{eq:pol}
 \xi^{\rm f}_1=\frac{ \xi_1 F_{11}}{F_0+\xi_3 F_3}, \quad
 \xi^{\rm f}_2=\frac{ \xi_2 F_{22}}{F_0+\xi_3 F_3}, \quad
 \xi^{\rm f}_3=\frac{F_3+ \xi_3 F_{33}}{F_0+\xi_3 F_3}.
\end{equation}
Note that the circular polarization of the secondary photon occurs only if the incident photon is circularly polarized.
The degree of polarization of the scattered photon is
\begin{equation}\label{eq:polfinal}
 \Pi=\sqrt{(\xi^{\rm f}_1)^2+(\xi^{\rm f}_2)^2+(\xi^{\rm f}_3)^2}.
\end{equation}
The polarization direction of the scattered photon can be expressed as
\begin{equation}\label{eq:direction}
  \tan2\chi=\frac{\xi_1^{\rm f}}{\xi_3^{\rm f}},
\end{equation}
where $\chi$ is the angle between the polarization direction and the $x'$-axis.

We consider that the incident photon beam is of synchrotron origin. Suppose that the magnetic field is globally uniform, the electrons are isotropic with a power-law distribution, i.e., $\mathcal{N}(\gamma)d\gamma\propto\gamma^{-p}d\gamma$. Then the synchrotron photons are linearly polarized, and the polarization degree is $\Pi_0= (p+1)/(p+7/3)$ \citep{Rybicki:1979}. The Stokes parameters of the synchrotron photons can be written as
\begin{align}\label{eq:xi0}
 \xi_1=&\Pi_0 \sin 2 \chi_0, \quad
 \xi_2=0, \quad
 \xi_3= \Pi_0 \cos 2 \chi_0.
\end{align}
After being scattered by electrons, the Stokes parameters change according to Eq.(\ref{eq:pol}), and the polarization direction changes according to Eq.(\ref{eq:direction}).

We aim to obtain the polarization when a beam of synchrotron photons are scattered by isotropic electrons with a power-law distribution. After averaging over the energy and angles of the incident electrons, we obtain the averaged cross section as
\begin{align}\label{eq:csa}
 \big\langle\frac{d\sigma}{d\Omega}(\varepsilon_0,\theta)\big\rangle=&\frac{1}{C}\int_{\gamma_1}^{\gamma_2}{\cal N(\gamma)}d\gamma\int_0^{\pi}\sin\theta_2d\theta_2\int_0^{2\pi}d\varphi_2\frac{d\sigma}{d\Omega}\nonumber\\
 \equiv &\frac{1}{4} r_e^2\bigg[\langle F_0 \rangle+ \langle F_3 \rangle (\xi_3+\xi'_3) +\langle F_{11} \rangle \xi_1 \xi_1'+\langle F_{22} \rangle \xi_2\xi'_2+\langle F_{33} \rangle \xi_3\xi'_3\bigg],
\end{align}
where $\gamma_1$ and $\gamma_2$ are the lower and upper boundary of the Lorentz factors of the incident electrons, respectively. $C\equiv\int_{\gamma_1}^{\gamma_2}{\cal N(\gamma)}d\gamma\int_0^{\pi}\sin\theta_2d\theta_2\int_0^{2\pi}d\varphi_2$ is the normalization factor. The scattered photons have an axial symmetry relative to the $z$-axis, so that $d\sigma$ is independent of the azimuthal angle $\varphi$. The definitions of the averaged components $\langle F_a \rangle$ ($a=0,3,11,22,33$) are similar to that of the averaged cross section, i.e.,
\begin{equation}\label{eq:avf}
 \langle F_a (\varepsilon_0,\theta)\rangle=\frac{1}{C}\int_{\gamma_1}^{\gamma_2}{\cal N(\gamma)} d\gamma\int_0^{\pi}\sin\theta_2 d\theta_2\int_0^{2\pi}d\varphi_2\left(\frac{\varepsilon_1}{\varepsilon_0}\right)^2 F_a.
\end{equation}
Here the $\varepsilon_1^2/\varepsilon_0^2$ term is included, because the Stokes parameters are the ratio of intensities, which are proportional to the cross section. The final Stokes parameters can be obtained by replacing the components $F_a$ in Eq.(\ref{eq:pol}) with the averaged components $\langle F_a \rangle$.

It is difficult to give an analytical expression of these components. Numerical method is applied to estimate the final polarization. After integrating over the parameter space of the incident electrons ($\gamma$, $\theta_2$, $\varphi_2$), one can obtain the degree of polarization $\Pi$ as a function of the viewing angle $\theta$ and the incident photon energy $\varepsilon_0$. In the cases of static electrons, power-law electrons and thermal electrons, the results are already given in the paper of \citet{Chang:2013yma}. It was showed that the scattered photons can be completely polarized if the electrons are at rest. Significant polarization can still be realized even if the electrons are isotropic, regardless that the electrons have thermal or non-thermal energy spectra.

All of the formulae above are only valid in the jet frame. In  order to make them convenient to be used, we should convert the quantities into the observer frame. On the other hand, the polarization as a function of the incident photon energy $\varepsilon_0$, should be converted to that of the scattered photon energy $\varepsilon_1$. To do this, note that the scattering angle between the two frames are related by \citep{Rybicki:1979}
\begin{equation}\label{eq:angle-transform}
  \cos\theta=\frac{\cos\bar{\theta}-B}{1-B\cos\bar{\theta}},
\end{equation}
where $B=(1-1/\Gamma^2)^{1/2}$ is the velocity of the jet moving towards the observer, and $\Gamma$ is the Lorentz factor of the jet. Here and after, symbols with a bar denote the quantities in the observer frame. From Eq.(\ref{eq:angle-transform}), we can see that, for $\theta$ not around $\pi$, $\bar{\theta}$ always approximates to zero. This means that isotropic electrons in the jet frame seem to be collimated along the line of jet direction in the observer frame. Besides, Eq.(\ref{eq:epspr}) can be equivalently converted to the form
\begin{equation}\label{eq:energy-convert}
  \frac{\varepsilon_0}{\varepsilon_1}=\frac{1-\beta\cos\theta_1}{(1-\beta\cos\theta_2)-\frac{\varepsilon_1}{\gamma m_ec^2}(1-\cos\theta)},
\end{equation}
where the energy of the scattered photon in the jet frame can be Doppler-shifted to that in the observer frame
\begin{equation}\label{eq:Doppler-shift}
  \varepsilon_1=\frac{\bar{\varepsilon}_1}{\Gamma(1+B\cos\theta)}.
\end{equation}
Substituting Eqs.(\ref{eq:angle-transform})(\ref{eq:Doppler-shift}) into Eq.(\ref{eq:energy-convert}), one obtains
\begin{equation}\label{eq:energy-convert2}
  \frac{\varepsilon_0}{\varepsilon_1}=\frac{1-\beta\cos\theta_1}{(1-\beta\cos\theta_2)-\frac{\Gamma\bar{\varepsilon}_1}{\gamma m_ec^2}(1+B)(1-\cos\bar{\theta})}.
\end{equation}

Note that the energies of both the incident and scattered photons are positive. Thus, the denominator of the right-hand-side of Eq.(\ref{eq:energy-convert2}) should be positive. This gives a constraint on the parameter space of electrons, i.e.,
\begin{equation}\label{eq:physical-range}
  \cos\theta_2<\frac{1}{\beta}\left(1-\frac{\Gamma\bar{\varepsilon}_1\bar{\theta}^2}{\gamma m_ec^2}\right)\equiv K_1.
\end{equation}
Here, we have used the approximation that $\Gamma\gg 1$ and $\bar{\theta}\ll 1$. The meaning of this constraint is that, for a given scattered photon with energy $\bar{\varepsilon}_1$ and observed at angle $\bar{\theta}$, the electrons which contribute to the scattering process must satisfy Eq.(\ref{eq:physical-range}). In other words, the electrons out of the constraint cannot scatter a photon, regardless of its initial energy, to angle $\bar{\theta}$ and energy $\bar{\varepsilon}_1$. Substituting Eqs.(\ref{eq:angle-transform})(\ref{eq:energy-convert2}) into Eqs.(\ref{eq:fs})\----(\ref{eq:xy}), and integrating over the parameter space of the incident electrons (constrained by Eq.(\ref{eq:physical-range})), we can obtain the averaged components $\langle F_a \rangle$. Substituting $\langle F_a \rangle$ into Eq.(\ref{eq:pol}), the Stokes parameters of the final photon can be derived.

\section{Special cases}\label{sec:result}

The numerical calculation in the most general case is time consuming. In this section, we will consider four special cases: (1) electrons are at rest in the jet frame, (2) Thomson limit $\varepsilon_0\ll m_ec^2$, (3) head-on collisions between beamed photons and beamed electrons, (4) electrons are monochromatic and isotropic. Although these four cases may be far away from the actual astrophysical processes, they can show the main properties of photon polarization via Compton scattering process.

\subsection{Electrons at rest}\label{sec:rest}

Firstly, we investigate the most simple case, in which the incident electrons are at rest in the jet frame. Such a system has already been studied in detail \citep{Chang:2013}. We will show that it is a special case of the setup of the present paper. Since electrons are static, one has $\beta=0$ and $\gamma=1$. For an initially linearly polarized photon, whose Stokes parameters are given in Eq.(\ref{eq:xi0}), the final Stokes parameters after scattering are reduced to
\begin{align} \label{eq:popost}
 \xi^{\rm f}_1=&\frac{2 \Pi_0 \sin 2\chi_0 \cos \theta}{\varepsilon_1/\varepsilon_0+\varepsilon_0/\varepsilon_1-(1-\Pi_0 \cos 2 \chi_0)\sin^2 \theta}, \nonumber \\
\xi^{\rm f}_3=&\frac{\sin^2 \theta+ \Pi_0 \cos 2\chi_0(1+\cos^2 \theta)}{\varepsilon_1/\varepsilon_0+\varepsilon_0/\varepsilon_1-(1-\Pi_0 \cos 2\chi_0)\sin^2 \theta},
\end{align}
where $\theta$ is related to $\bar{\theta}$ through Eq.(\ref{eq:angle-transform}), and $\varepsilon_1/\varepsilon_0$ is derived from Eq.(\ref{eq:energy-convert2}) by setting $\gamma=1$ and $\beta=0$, i.e.,
\begin{equation}\label{eq:energy-convert-rest}
  \frac{\varepsilon_1}{\varepsilon_0}=1-\frac{\Gamma\bar{\varepsilon}_1}{m_ec^2}(1+B)(1-\cos\bar{\theta}).
\end{equation}
The photon observed at angle $\bar{\theta}$ has an energy cutoff $\bar{\varepsilon}_1<m_ec^2/\Gamma\bar{\theta}^2$. When the polarization direction of the incident photon is parallel to the scattering plane, i.e., $\chi_0=\pi/2$, we have $\xi^{\rm f}_1=0$ and $\Pi=-\xi^{\rm f}_3$. This agrees with the result given by \citet{Chang:2013}. As was pointed out, an interesting feature of the polarization in such a setup is that the direction of polarization may change $90^{\circ}$ after the scattering process. If the incident photon is unpolarized and its energy is low enough such that the Thomson limit is valid, i.e., $\Pi_0=0$ and $\varepsilon_1=\varepsilon_0$, Eq.(\ref{eq:popost}) reduces to the famous result $\Pi=-\xi_3^{\rm f}=-(1-\cos^2\theta)/(1+\cos^2\theta)$. The positive $\xi_3$ means that the polarization direction of the scattered photon is always perpendicular to the scattering plane.

\begin{figure}
\centering
  \plotone{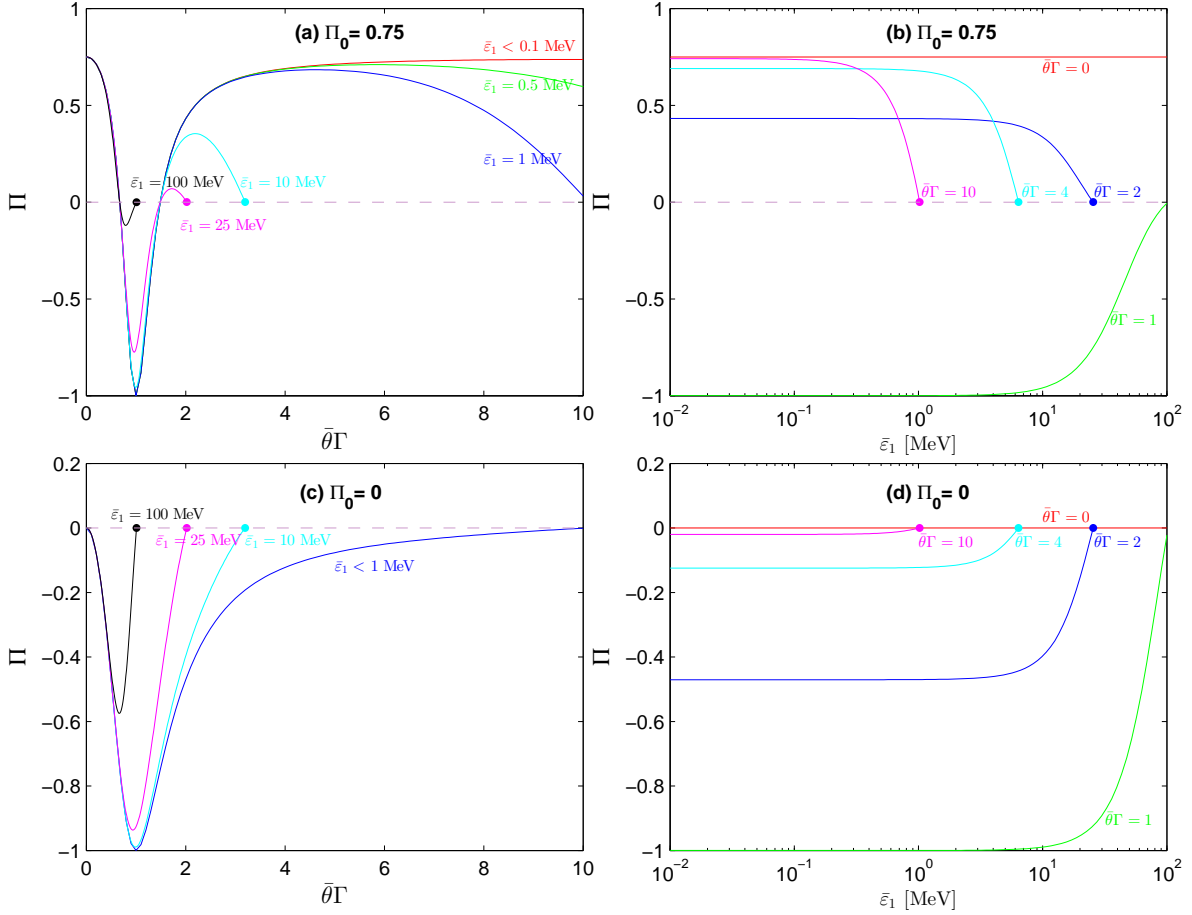}
  \caption{\small{The polarization of photons scattered by electrons at rest. The upper two panels depict that the incident photons have an initial polarization $\Pi_0=0.75$, while the lower two panels depict that the incident photons are completely unpolarized. The dot on the end of each curve stands the energy cutoff. We set $\chi_0=\pi/2$ and $\Gamma=200$ in the numerical calculation.}} \label{fig:Electron-at-rest}
\end{figure}

We plot the polarization as a function of photon energy $\bar{\varepsilon}_1$ and viewing angle $\bar{\theta}$ in Fig.\ref{fig:Electron-at-rest}. The upper two panels depict that the incident photons have an initial polarization $\Pi_0=0.75$, while the lower two panels depict that the incident photons are completely unpolarized. The dot (if there has) on the end of each curve stands the energy cutoff \footnote{Curves without a dot mean that the energy cutoff is beyond the plot range.}. We set $\chi_0=\pi/2$ in the numerical calculation. The Lorentz factor of the jet is taken to be $\Gamma=200$, the typical value of long GRBs \citep{Chang:2012}. The positive and negative values of $\Pi$ stand for the polarization parallel and perpendicular to the scattering plane, respectively. From Fig.\ref{fig:Electron-at-rest}, we can see that high polarization can be achieved even if the incident photons are unpolarized. High energy photons have small polarization than low energy photons. The low energy photons can be completely polarized at $\bar{\theta}\Gamma\approx 1$. At a fixed viewing angle, photons with energy $\bar{\varepsilon}_1\lesssim 0.1$ MeV almost have the same polarization. The most interesting feature is that, for the initially polarized photons, the polarization direction can change $90^{\circ}$ after scattering at certain viewing angles. Note that the change of polarization angle occurs only if the incident photon is polarized. In the initially unpolarized case, we can see from panel (c) of Fig.\ref{fig:Electron-at-rest} that $\Pi$ is always negative (equivalently, $\xi_3^{\rm f}$ is always positive). Thus, the polarization direction of scattered photon is always perpendicular to the scattering plane. This is a well-known result of Thomson scattering. If the incident photon is partially polarized, see panel (a) of Fig.\ref{fig:Electron-at-rest} for example, the sign of $\xi_3^{\rm f}$ can be changed from negative to positive (equivalently, the sign of $\Pi$ changed from positive to negative) at $\bar{\theta}\Gamma\sim 1$. Thus, as the viewing angle increases, the $90^{\circ}$ change of polarization angle can be observed.

For photons with low energy $\bar{\varepsilon}_1\lesssim 0.1$ MeV, the polarization reaches its maximum $\Pi_{\rm max}\approx 1$ at the viewing angle $\bar{\theta}\Gamma\approx 1$. This can been easily understood. It is well known that in the Thomson limit, the polarization of an initially unpolarized photon scattered by an electron at rest is $\Pi(\theta)=(1-\cos^2\theta)/(1+\cos^2\theta)$ \citep{Rybicki:1979}. The scattered photon is completely polarized at $\theta=\pi/2$. When transform the angle to the observer frame, we have $\bar{\theta}\approx 1/\Gamma$. This is because that, from Eq.(\ref{eq:angle-transform}), we get $\cos\bar{\theta}=(\cos\theta+B)/(1+B\cos\theta)\mid_{\theta=\pi/2}=B=(1-1/\Gamma^2)^{1/2}\approx 1-1/2\Gamma^2\approx \cos(1/\Gamma)$. Here we have assumed that $\Gamma\gg 1$.

\subsection{Thomson limit}\label{sec:Thomson-limit}

The photon polarization via Compton scattering process has been well studied in the Thomson limit \citep{Bonometto:1970,Bonometto:1973}. When the energy of the incident photon is much smaller than the rest mass energy of electron, i.e., $\varepsilon_0\ll m_ec^2$, the Thomson limit is valid. In this case, the formulae in section \ref{sec:general-formulae} can be highly simplified. Eq.(\ref{eq:epspr}) reduces to
\begin{equation}
\varepsilon_1=\frac{\varepsilon_0(1-\beta \cos\theta_2)}{1-\beta\cos\theta_1}.
\end{equation}
Then from Eqs.(\ref{eq:ab})(\ref{eq:xy}), we can see that $x=y$, $A=0$ and $B=2$.  Thus, Eq.(\ref{eq:fs}) becomes
\begin{equation}
  F_0=\frac{\varepsilon_1}{\varepsilon_0}+\frac{\varepsilon_0}{\varepsilon_1}-\sin^2\theta, \quad F_3=0, \quad F_{11}=F_{22}=F_{33}=\frac{1}{2}\Sigma,
\end{equation}
where $\Sigma$ is given by Eq.(\ref{eq:ab}). Note that in this case, all of the components $F_a$ are independent of photon energy. Thus, the final polarization is also energy independent. The Stokes parameters of the scattered photons simplify to $\xi_1^{\rm f}=\xi_1F_{11}/F_0$, and $\xi_3^{\rm f}=\xi_3F_{33}/F_0$. After averaging over the parameter space of the incident electrons, the polarization of the scattered photons reads
\begin{equation}\label{eq:pi-thomson}
  \Pi=\Pi_0\frac{\langle F_{11} \rangle}{\langle F_0 \rangle}.
\end{equation}
Eq.(\ref{eq:pi-thomson}) shows that the final polarization is proportional to the initial polarization of the incident photons. Unpolarized photons are still unpolarized after scattering. The final polarization is independent of the polarization direction of the incident photons.

\begin{figure}
\centering
  \plotone{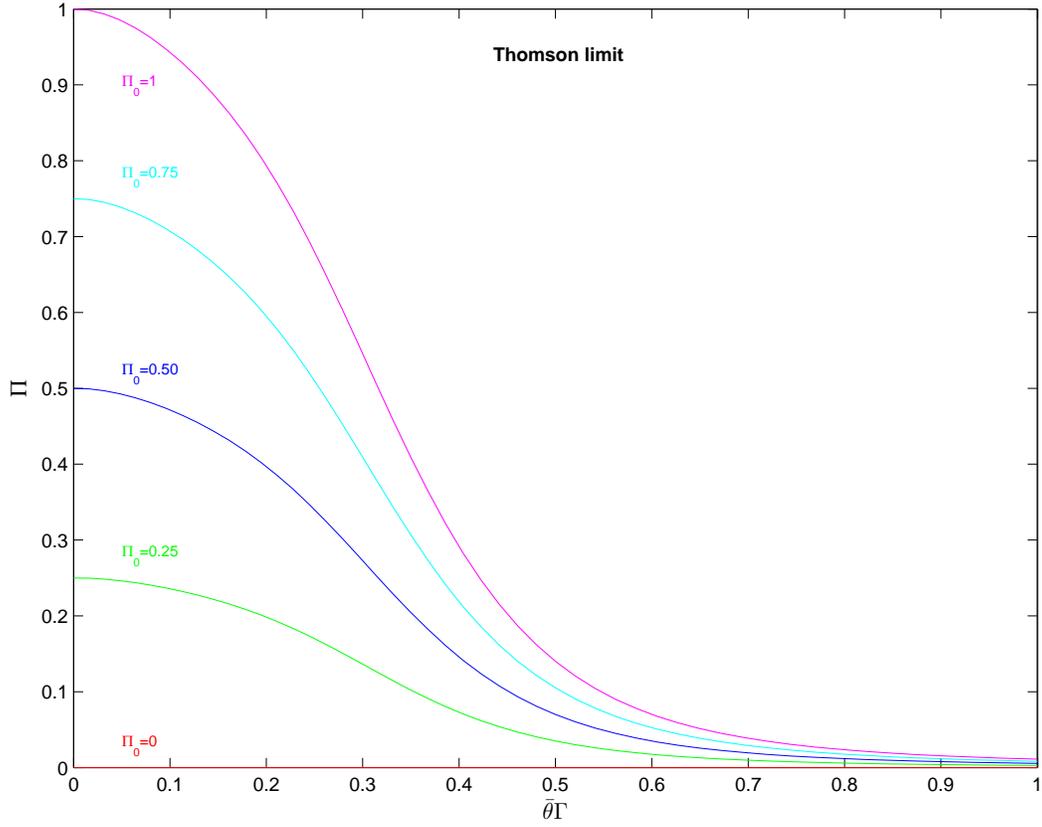}
  \caption{\small{The polarization of low energy (Thomson limit is valid) photons scattered by isotropic electrons in power-law distribution. The polarization as a function of viewing angle is plotted for various initial polarization. We take $\Gamma=200$ in the numerical calculation.}}\label{fig:Thomson-limit}
\end{figure}

In the case of electrons are isotropic with a power-law distribution, i.e., ${\cal N}(\gamma)d\gamma\propto \gamma^{-p}d\gamma$, we plot the polarization of the scattered photons as a function of the viewing angle in Figure \ref{fig:Thomson-limit}. We take $p=3$ as a typical value in the numerical calculation. The Lorentz factor of the electrons is taken to be in the range $\gamma\in[1,10]$. Electrons with Lorentz factor larger than 10 almost have no contribution to the polarization \citep{Chang:2013yma}. The Lorentz factor of the jet is assumed to be $\Gamma=200$. Different curves stand for different initial polarization. As can be seen, the polarization decreases monotonically as the viewing angle increases. At $\bar{\theta}\Gamma=0$, when the photons are not scattered, the polarization of the final photons equals its initial value. When $\bar{\theta}\Gamma$ approaches 1, the polarization almost vanishes. Another noticeable feature is that, an unpolarized beam is still unpolarized after scattering. \citet{Bonometto:1970} obtained the analytical formulae of polarization of the Compton scattering for arbitrary distributions of photons and electrons in the Thomson limit. In the special case when a beam of unpolarized photons scattered by isotropic electrons, they also found that the scattered photons are still unpolarized. Our conclusion is consistent with that of \citet{Bonometto:1970}. But the general formulae in this paper are still valid in the Klein\---Nishina region.

\subsection{Head-on collision}\label{sec:head-on}

We consider the head-on collision between a beam of photons and a beam of electrons. The energy of electrons is assumed to be in power-law distribution. In this case, one has
\begin{equation}\label{eq:head-on}
  \theta_2=\pi, \quad \varphi_2=0, \quad \theta_1=\pi-\theta.
\end{equation}
Thus, Eq.(\ref{eq:ab}) simplifies to
\begin{equation}
  A=-\frac{1-\beta}{1+\beta\cos\theta}\sin^2\frac{\theta}{2}, \quad
  B=\frac{\varepsilon_0(1+\beta)}{\varepsilon_1(1+\beta\cos\theta)}+\frac{\varepsilon_1(1+\beta\cos\theta)}{\varepsilon_0(1+\beta)}, \quad
  \Sigma=4\frac{1-\beta}{1+\beta},
\end{equation}
where $\varepsilon_0/\varepsilon_1$ is derived by substituting Eq.(\ref{eq:head-on}) in to Eq.(\ref{eq:energy-convert2}), i.e.,
\begin{equation}
  \frac{\varepsilon_0}{\varepsilon_1}=\frac{1+\beta\cos\theta}{(1+\beta)-\frac{\Gamma\bar{\varepsilon}_1}{\gamma m_ec^2}(1+B)(1-\cos\bar{\theta})},
\end{equation}
and $\cos\theta$ is given by Eq.(\ref{eq:angle-transform}). The constraint on the parameter space of the incident electrons Eq.(\ref{eq:physical-range}) becomes
\begin{equation}\label{eq:physical-range2}
  \gamma+\sqrt{\gamma^2-1}>\frac{\Gamma\bar{\varepsilon}_1\bar{\theta}^2}{m_ec^2}\equiv K_2.
\end{equation}
If the Lorentz factors of the incident electrons are in the range $\gamma\in[1,10]$, then the solution of Eq.(\ref{eq:physical-range2}) is
\begin{eqnarray}
  \begin{cases}
    1\leq\gamma\leq10 \quad & {\rm for} ~~ K_2<1, \\
    \frac{1+K_2^2}{2K_2}<\gamma<10 \quad & {\rm for} ~~ K_2\geq 1~{\rm and}~\frac{1+K_2^2}{2K_2}<10, \\
    {\rm no~solution} \quad & {\rm for} ~~ \frac{1+K_2^2}{2K_2}\geq 10.
  \end{cases}
\end{eqnarray}

\begin{figure}
\centering
  \plotone{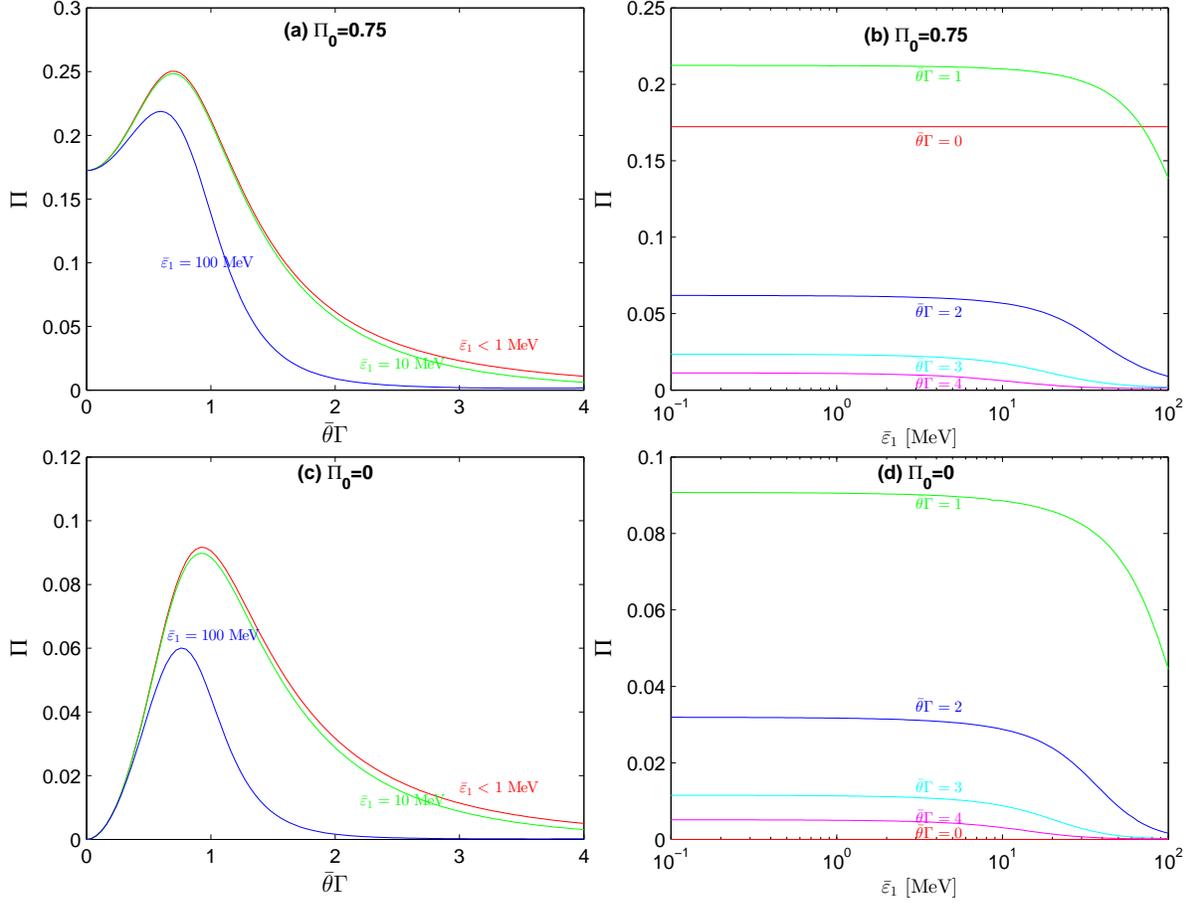}
  \caption{\small{The polarization of photons after head-on collision with a beam of electrons in power-law distribution. The upper two panels depict that the incident photons have an initial polarization $\Pi_0=0.75$, while the lower two panels depict that the incident photons are completely unpolarized. We set $\chi_0=0$ and $\Gamma=200$ in the numerical calculation.}} \label{fig:Head-on}
\end{figure}

The final polarization as a function of $\bar{\varepsilon}_1$ and $\bar{\theta}$ is plotted in Fig.\ref{fig:Head-on}. The upper two panels depict that the incident photons have an initial polarization $\Pi_0=0.75$, while the lower two panels depict that the incident photons are completely unpolarized. We set $\chi_0=0$ in the numerical calculation. The Lorentz factors of the jet is taken to be $\Gamma=200$. The spectrum index of the electrons is chosen to be $p=3$. As can be seen, for the low energy photons $\bar{\varepsilon}_1\lesssim 1$ MeV, the polarization reaches its maximum at $\bar{\theta}\Gamma\approx 1$. If the incident photons have an initial polarization $\Pi_0=0.75$, the polarization of the scattered photons can be as large as $25\%$. For the initially unpolarized photons, the final polarization is less than $10\%$. At the fixed viewing angle, high energy photons have smaller polarization than low energy ones. When $\bar{\theta}\Gamma\gtrsim 4$, the polarization almost vanishes.

A remarkable feature is that the final polarization differs from the initial polarization even at $\bar{\theta}\Gamma=0$. Consider the most general case when a beam of photons scattered by isotropic electrons. At the viewing angle $\theta=0$, where the photons are not scattered, we can see easily from Eqs.(\ref{eq:epspr})(\ref{eq:angrel}) that $\cos\theta_1=\cos\theta_2$ and $\varepsilon_1=\varepsilon_2$. This means that the momenta and energies of the photons are not changed after scattering. Then from Eqs.(\ref{eq:fs})\---(\ref{eq:xy}), we have $F_0=2$, $F_3=0$ and $F_{11}=F_{22}=F_{33}=\frac{1}{2}\Sigma$, where $\Sigma=4\gamma^{-2}(1-\beta\cos\theta_2)^{-2}$. The polarization of the scattered photons simplifies to $\Pi=\Pi_0\langle F_{11}\rangle /\langle F_0\rangle$, where the averaged components $\langle F_a\rangle$  are defied by Eq.(\ref{eq:avf}). After integrating over the parameter space of the incident electrons $(\gamma,\theta_2,\varphi_2)$, we find that $\langle F_{11}\rangle\equiv\langle F_0\rangle$. Thus, the finally polarization exactly equals initial polarization. However, if the incident electrons are beamed, the net polarization cannot vanish. This implies that the Compton scattering can change the polarization of a photon, but keeps its momentum and energy unchanged.

\subsection{Monochromatic electrons}\label{sec:monochromatic}

Consider the case when the incident electrons are isotropic and monochromatic, i.e., the Lorentz factor of the electrons is a constant. The solution of Eq.(\ref{eq:physical-range}) is
\begin{eqnarray}
  \begin{cases}
    0\leq\theta_2\leq\pi \quad & {\rm for}~~K_1> 1,\\
    \arccos(K_1)<\theta_2<\pi \quad &{\rm for}~~-1<K_1\leq 1,\\
    {\rm no~solution} \quad & {\rm for}~~K_1\leq -1.
  \end{cases}
\end{eqnarray}
The azimuthal angles of the electrons are in the range $\varphi_2\in[0,2\pi]$. After integrating over the parameter space of the incident electrons $(\theta_2, \varphi_2)$, we can obtain the polarization of the scattered photons as a function of $\bar{\varepsilon}_1$ and $\bar{\theta}$.

\begin{figure}
\centering
  \plotone{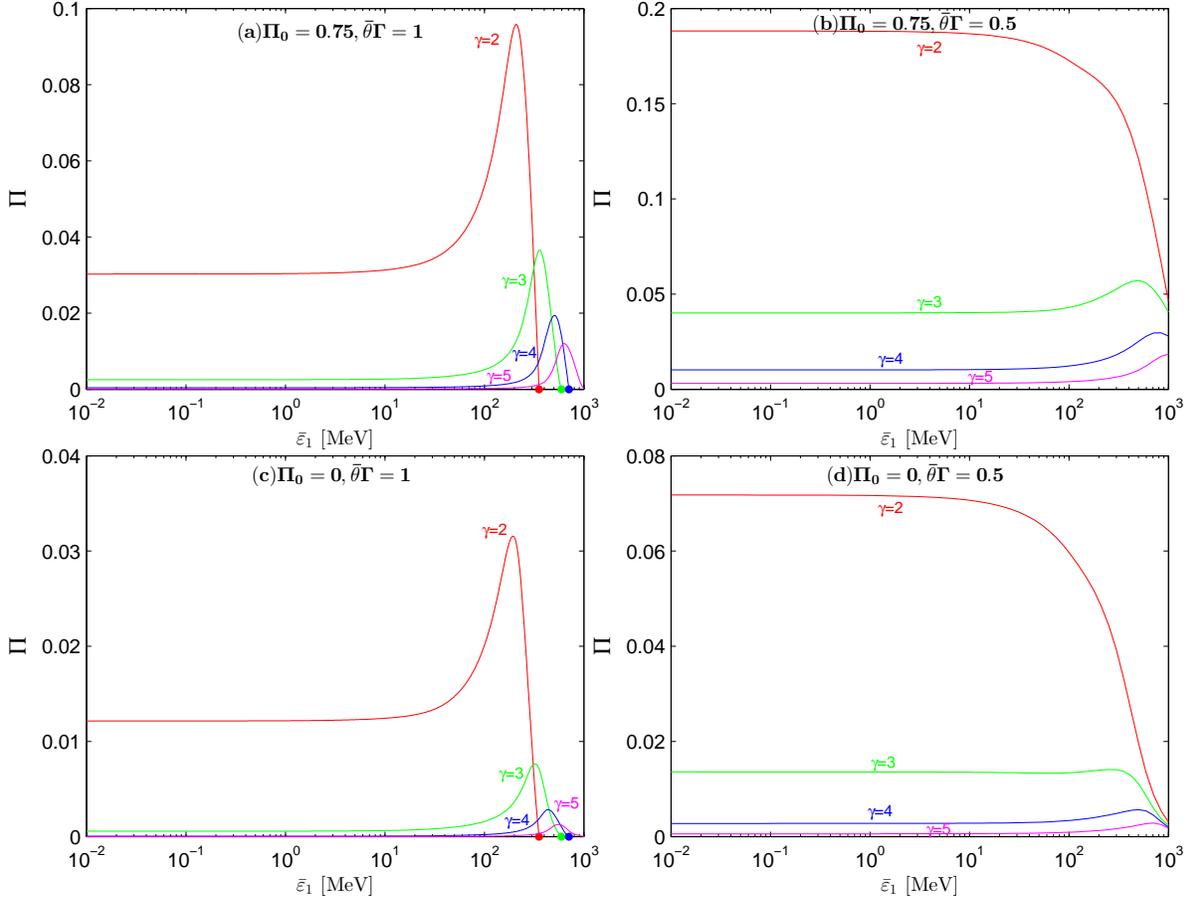}
  \caption{\small{The polarization of photons scattered by isotropic and monochromatic electrons. The upper two panels depict that the incident photons have an initial polarization $\Pi_0=0.75$, while the lower two panels depict that the incident photons are completely unpolarized. The left two panels represent the polarization observed at $\bar{\theta}\Gamma=1$, while the right two panels represent the polarization observed at $\bar{\theta}\Gamma=0.5$. The dot on the end of each curve stands for the energy cutoff. Curves for different $\gamma$ are plotted. We set $\chi_0=0$ and $\Gamma=200$ in the numerical calculation.}} \label{fig:monochromatic}
\end{figure}

We plot the polarization as a function of photon energy ${\bar\varepsilon}_1$ in Fig.\ref{fig:monochromatic}. The upper two panels depict that the incident photons have an initial polarization $\Pi_0=0.75$, while the lower two panels depict that the incident photons are completely unpolarized. The left two panels represent the polarization observed at $\bar{\theta}\Gamma=1$, while the right two panels represent the polarization observed at $\bar{\theta}\Gamma=0.5$. Curves for different $\gamma$ are plotted. We set $\chi_0=0$ and $\Gamma=200$ in the numerical calculation. From Fig.\ref{fig:monochromatic}, we can see that the polarization decreases fast as $\gamma$ increases. For $\gamma\gtrsim 5$, the polarization almost vanishes. This further indicates that the polarization effect mainly comes from the cold electrons. In the case when the initial photons have polarization $\Pi_0=0.75$, the final photons have maximum polarization $\Pi\approx 19\%$ at $\bar{\theta}\Gamma=0.5$ when $\gamma=2$. If the incident photons are initially unpolarized, the scattered photons have maximum polarization $\Pi\approx 7\%$ at $\bar{\theta}\Gamma=0.5$ when $\gamma=2$. An interesting feature is that the polarization seems to have a peak at $\bar{\varepsilon}_1/\Gamma\sim 1$ MeV. This is consistent with the result of \citet{Chang:2013yma}, in which the authors showed that the polarization reaches its maximum at $\varepsilon_1\sim 1$ MeV in the jet frame. At the specific viewing angle $\bar{\theta}\Gamma=1$, we can see from panel (a) and panel (c) of Fig.\ref{fig:monochromatic} that the polarization almost vanishes at the low-energy end (where the Thomson limit is valid). This is consistent with the results of section \ref{sec:Thomson-limit}.

\section{Against observational data}\label{sec:GRB-cases}

We have obtained the numerical results of photon polarization via Compton scattering process above. In this section, we test the validity of the results in explaining the observational data on specific GRBs. Polarization has been observed in many GRBs in the prompt phase, such as GRB 021206 \citep{Rutledge:2004,Wigger:2004}, GRB 041219A \citep{McGlynn:2007,Kalemci:2007}, GRB 100826A \citep{Yonetoku:2011}, GRB110301A and GRB 110721A \citep{Yonetoku:2012}. Here we mainly concentrate on two GRBs, i.e., GRB 041219A and GRB 100826A, both of which are among the brightest GRBs ever observed at present. The polarization of these two GRBs shows some interesting features.

GRB 041219A is one of the longest and brightest GRBs ever observed at present time. It was detected by {\it INTEGRAL} at 01:42:18 UT on December 19th 2004 \citep{{Gotz:2004}}. The spectrum of GRB 041219A in the most intense pulse of duration 66 seconds can be well fitted by the Band function, with the spectrum indexes $\alpha=-1.50_{-0.06}^{+0.08}$ and $\beta=-1.95_{-0.21}^{+0.08}$ \citep{McGlynn:2007}. A search for linear polarization during the 66 seconds was performed in the energy bands $100\--350$ keV, $100\--500$ keV and $100\--1000$ keV. The polarization degrees are $63_{-30}^{+31}\%$, $49\%\pm 24\%$ and $26\%\pm 20\%$, respectively \citep{McGlynn:2007}. Furthermore, the polarization degrees during the brightest 12 s are $96_{-40}^{+39}\%$, $70\%\pm 37\%$ and $68\%\pm 29\%$ in the three energy bands, respectively. No significant polarization angle change was observed. These results imply that the polarization of high energy photons is smaller than that of the low energy ones. An independent analysis of the same burst shows that the polarization is reduced if high energy photons are included \citep{Kalemci:2007}. This tendency is coincident with the prediction of Compton scattering process. For isotropic electrons with power-law index $p$ moving in uniform magnetic field, the synchrotron theory predicts that the maximum polarization is $\Pi=(p+1)/(p+7/3)$ \citep{Rybicki:1979}, which is independent of photon energy. For a typical value $p\approx 3$, the maximum polarization is about $75\%$. If the magnetic field is nonuniform, the polarization is much smaller. The high and energy-dependent polarization of GRB 041219A implies that it is unlikely to be of synchrotron origin. The polarization effect provides a perfect tool to distinguish the radiation mechanisms of astrophysical processes such as GRBs.

GRB 100826A is one of the top $1\%$ brightest bursts listed in the {\it BATSE} catalog. It was detected by {\it IKAROS-GAP} at 22:57:20.8 UT on August 26th 2010. The spectrum of GRB 100826A can be well fitted by the Band function, with the spectrum indexes $\alpha=-1.31_{-0.05}^{+0.06}$, $\beta=-2.1_{-0.2}^{+0.1}$, and $\nu F_{\nu}$ peak energy $E_p=606_{-109}^{+134}$ keV \citep{Golenetskii:2010}. \citet{Yonetoku:2011} analyzed the data of the prompt emission of duration 100 seconds, and found a change of polarization angle  with $99.9\%$ confidence level, and the average polarization degree is $27\%\pm 11\%$ with $99.4\%$ confidence level. They divided the total duration (100 s) into two time intervals, labeled Interval-1 (47 s) and Interval-2 (53 s), respectively. The polarization degree and polarization angle during these two intervals are $\Pi_1=25\%\pm 15\%$, $\phi_1=159^{\circ}\pm 18^{\circ}$ and $\Pi_2=31\%\pm 21\%$, $\phi_2=75^{\circ}\pm 20^{\circ}$, respectively. There is high probability that the polarization changes $\sim 90^{\circ}$ between the two time intervals. \citet{Yonetoku:2011} pointed out that it is difficult to explain the observed significant change of polarization angle within the framework of axisymmetric jet. This phenomenon can be explained naturally by the synchrotron plus Compton scattering model considered in this paper. During the first interval, the viewing angle $\bar{\theta}$ is much smaller than the jet opening angle $\theta_{\rm jet}$, and the polarization is positive, which means that the polarization direction is parallel to the scattering plane. As time goes on, the jet spreads transversely, and the line-of-sight moves away from the jet axis. At a certain angle, the polarization changes from positive to negative, and the polarization direction becomes perpendicular to the scattering plane.

\section{Discussions and conclusions}\label{sec:conclusions}

In this paper, we obtained the analytical formulae for the polarization of beamed photons scattered by isotropic electrons with a power-law distribution\footnote{The formulae in this paper are also valid to other electron distributions, such as thermal distribution. The only thing one should do is to replace the power-law distribution $\mathcal{N(\gamma)}d\gamma\propto\gamma^{-p}d\gamma$ with other distributions. We just take the power-law distribution as an example in the numerical calculation.}. The incident photons are assumed to originate from synchrotron radiation, with any initial polarization degree $\Pi_0$ and polarization angle $\chi_0$. After being scattered by electrons, both the polarization degree and polarization angle are changed. The polarization of the scattered photons is a function of photon energy $\bar{\varepsilon}_1$ and viewing angle $\bar{\theta}$. In four special cases, we carried out numerical calculations. Although these four special cases may be far away from the actual astrophysical processes, they show the main features of photon polarization via Compton scattering process:
\begin{itemize}
\item{The Compton process can produce a wide range of polarization, from completely unpolarized to completely polarized.}
\item{At a fixed viewing angle, the polarization of high energy photons is smaller than that of the low energy photons.}
\item{The polarization effect mainly comes from the cold electrons. Electrons with Lorentz factor larger than 10 almost have no contribution to polarization.}
\item{In the electron rest case, low energy photons can be completely polarized at the viewing angle $\bar{\theta}\Gamma\sim 1$, while high energy photons cannot be completely polarized.}
\item{The polarization direction may be changed after scattering. In a special setup, the polarization direction can be changed $90^{\circ}$ exactly.}
\item{Due to the isotropic distribution of the electrons, the polarization is highly suppressed, but a maximum value of $\sim 10\%\--20\%$ can still be achieved.}
\item{If the electrons are isotropic and monochromatic, the polarization of the scattered photons has a peak at $\bar{\varepsilon}_1/\Gamma\sim 1$ MeV.}
\item{In the Thomson limit, i.e., $\varepsilon_0 \ll m_ec^2$, the polarization is independent of photon energy, and a beam of unpolarized photons are still unpolarized after scattering.}
\end{itemize}

The polarization properties of Compton scattering process are very different from that of other radiation mechanisms, such as synchrotron radiation and thermal radiation. The polarization measurement provides an excellent tool to distinguish different radiation mechanisms and jet structures. The observation of GRB 041219A shows that the polarization tends to decrease as the photon energy increases. This is consistent with the prediction of the Compton scattering process. The high and energy-dependent polarization of GRB 041219A implies that it is not likely to be of synchrotron origin. The maximum polarization of synchrotron radiation is $\sim 70\% \-- 80\%$. If the magnetic field is nonuniform, the polarization is much smaller. Furthermore, the Compton process can naturally explain the $90^{\circ}$ change of polarization angle observed in GRB 100826A, which is a challenge to most other models.  \citet{Yonetoku:2011} argued that the jet should be non-axisymmetric in order to have a significant change of polarization angle. However, they cannot explain why the polarization angle changes $90^{\circ}$ exactly.

Although polarization has been observed in many GRBs, the observational energy band is usually limited to be below 1 MeV. The polarization measurement in high energy band is still lacking.  Besides, the observation is often limited in a narrow energy band. Observing the polarization in different energy bands is desirable. GRB 100826A, to our knowledge, is among the few bursts which have a significant change of polarization angle. A more detailed analysis of GRB 041219A in different time intervals shows that the polarization degree and polarization angle change dramatically with time \citep{Gotz:2009}. The time-averaged value over longer intervals shows reduced polarization. The variation of jet Lorentz factor may lead to the variation of polarization. According to the calculation in the above sections, the change of polarization angle is common in Compton scattering process. The future measurement of polarization in energy bands higher than 1 MeV and the change of polarization angles may provide us deeper insight into the radiation mechanism and the structure of emission region.

The geometry considered in this paper is over-simplified. We have assumed that the photons are scattered only once by electrons. The actual process is much more complex. In the optically thick region, multiple scattering occurs. After one scattering process, a photon changes its polarization as well as momentum and energy, and beamed photons become unbeamed. For such a multi-scattering system, it is very difficult to give analytical formulae to calculate polarization. Since for unbeamed photons, there does not exist a coordinate system which is suitable to describe the Stokes parameters for all the photons. The numerical simulation is useful to calculate polarization in such a complex system. In fact, Monte Carlo simulation has already been used to calculate polarization in the multi-scattering case \citep{McNamara:2009,Krawczynski:2012}. The formulae in this paper are easy to be used in Monte Carlo simulation, since they are expressed in the observer frame and can avoid Lorentz transformation between different frames. On the other hand, if the energy of photon is larger than the rest mass energy of electron, the $\gamma+\gamma\rightarrow e^++e^-$ annihilation may also take place. This process may significantly affect the polarization.

\begin{acknowledgments}
We are grateful to X. Li, S. Wang and D. Zhao for useful discussion. The work of Z. Chang and H.-N. Lin has been funded by the National Natural Science Fund of China under Grant No. 11075166 and No. 11375203. The work of Y.~G.~Jiang has been funded by NSFC under Grant No. 11203016 and No. 11143012.
\end{acknowledgments}

\end{document}